\documentstyle[editedvolume,numreferences]{crckapb}

\input{epsf}

\newcommand{\keyword}[1]{\index{#1}#1}

\begin{opening}
\title{Constrained Path Monte Carlo For Fermions}

\author{SHIWEI ZHANG}
\institute{Department of Physics and Department of Applied Science\\
           College of William \& Mary, Williamsburg, VA 23187, USA}

\end{opening}

\runningtitle{Constrained path Monte Carlo}

\makeindex

\begin{document}

\keyword{CPMC}

\setcounter{table}{1}

\section{Introduction}

In these lectures we describe the constrained path Monte Carlo (CPMC)
method for quantum many-fermion systems. We will focus on the
ground-state CPMC algorithm\cite{cpmc0K} and its
applications\cite{pairing}, but will also briefly describe work in
progress on a finite-temperature extension\cite{finiteT}.

In previous lectures we have seen both the auxiliary-field quantum
Monte Carlo (AFQMC)\cite{BSS,AFQMC} and Green's function Monte Carlo
(GFMC) \cite{GFMC} methods. We will see here how the ground-state
fermion CPMC algorithm combines the concepts of Hubbard-Stratonovich
transformation and Slater determinants with branching random walks and
importance sampling. We will demonstrate the origin of the fermion
sign problem\cite{sign1,sign2} in this context and then discuss an
approximate solution, the constrained path approximation. CPMC is free
of any sign decay and its computing time scales algebraically with
system size.  The Hubbard model will be used as an example to
illustrate the basic idea and actual implementation of the algorithm,
although the formalism of the algorithm is more general.

Having introduced the Hubbard model and discussed the ground-state
CPMC method, we will then describe in some detail an application of
CPMC. The Hubbard model has been the subject of intense theoretical
effort for the past decade, to study its ground-state properties and
understand whether it contains the relevant electron correlations to
describe high-Tc superconductivity\cite{Dagotto,scalapino95}. CPMC
enabled, for the first time, ground-state calculations on large system
sizes. On the other hand, the task of characterizing the electron
pairing correlations is an extremely difficult one and requires great
accuracy. This application illustrates both the promise and the
challenge facing our current fermion methods.

In the last part we will briefly discuss some preliminary results on a
finite-temperature ($T>0\,$K) CPMC method. The goal is to have an
algorithm which enables calculations at finite temperatures free of
any sign decay, while maintaining much of the grand-canonical
formalism of the standard Blankenbecler, Scalapino, and Sugar (BSS)
algorithm\cite{BSS}.

Why do we need another approximate fermion method?  After all, we have
fixed-node\cite{Anderson,Moskowitz-Reynolds} Green's function Monte Carlo (GFMC) 
(including diffusion
Monte Carlo (DMC)\cite{Anderson,qmc_book}) and 
projector quantum Monte Carlo\cite{AFQMC,sorella}, for ground states, and
path integral Monte Carlo\cite{CepRMP,restr_path} and auxiliary-field 
quantum Monte Carlo\cite{BSS}
for finite-temperature properties. The short answer is that
there are {\em many\/} applications that demand it. In fact the
ground-state CPMC algorithm has already made possible various
calculations and is seeing an increasing number of applications in
different areas. We will also give a long answer to this question by
making a rough table of the existing algorithms, as shown 
on the next page.  We {\em loosely\/} categorize these fermion
algorithms into {\bf continuum} and {\bf lattice} methods.  The table
is a synopsis to capture some basic ideas, which will provide a very
brief review of some of the materials we have learned, and also help
set the reference frame for the algorithms we will discuss.  (Readers
unfamiliar with AFQMC may wish to first read Section 2.1.)

\begin{figure}
\vspace{-0.75in}
\epsfxsize=6.in
\epsfysize=9.6in
\centerline{\epsfbox{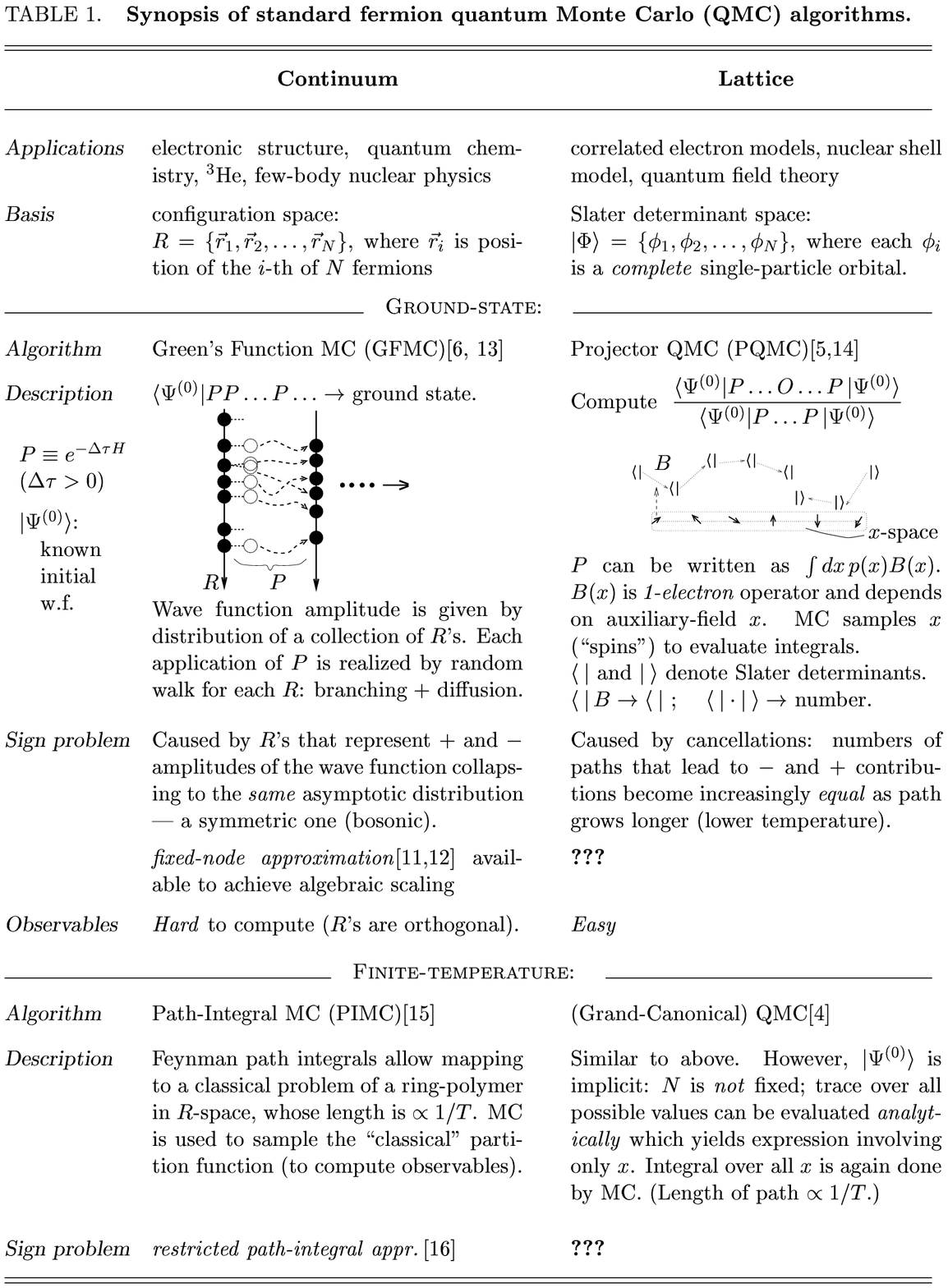}}
\end{figure}

While cross-fertilization certainly has occurred, the
division line between the two classes of methods remains quite
visible.  Each has distinct strengths, which have to a large extent
maintained their separation in application areas.  Although each has
seen a great deal of success, limitations do exist in both, as we
see from the table. In this summer institute, we have seen
considerably more ``mixing'' of the two, and we can hope that this
will stimulate even more such efforts.

\section{Ground-state ($T=0\,$K) CPMC algorithm}

The ground-state constrained path Monte Carlo (CPMC) algorithm filled
in the void (indicated by the first set of question marks in 
Table 1.)
of a ground-state auxiliary-field algorithm without exponential
scaling. CPMC is free of any ``sign decay'', which is the signature of
the fermion sign problem. The required computation time therefore
scales algebraically with system size. It is approximate, with results
dependent on the trial wave function that is used as constraint.
Similar to the fixed-node approximation \cite{Anderson,Moskowitz-Reynolds}
in configuration space, the
process of CPMC solves the Schrodinger equation under some boundary
condition defined by the trial wave function.  But, differently from
fixed-node, this boundary condition is not limited to real space, and
can be a more ``global'' one, as we will discuss in detail below. CPMC
also retains (at least to a large degree) an important advantage of
AFQMC compared to continuum methods, namely the ability to compute
easily many expectation values, including off-diagonal ones such as
electron pairing correlations. This is an important point since in
many applications, particularly those in the study of strongly
correlated models, it is crucial to have information on properties
beyond the total energy.

The ground-state CPMC algorithm has two main components: The {\it
first\/} is to formulate the projection of the ground state as
open-ended random walks with importance sampling, as in GFMC, although
the random walks take place in a space of Slater determinants. The
{\it second\/} is to constrain the paths of the random walks so that
any Slater determinant generated maintains a positive overlap with a
known trial wave function $|\psi_T\rangle$. Below we describe them in
sections 2.2 and 2.3 respectively. The first is an alternative way to
do projector QMC which in many cases can be more efficient than
the latter, because of the use of importance sampling. Like
projector QMC, it is exact, but suffers from the sign problem. The
second component is the constrained path approximation, which
eliminates the sign decay, but introduces a systematic error in the
algorithm. These two components are independent of each other, and can
be used separately. Their combination is what we are calling the
ground-state CPMC algorithm.

In describing the CPMC algorithm, we will use the Hubbard model
and often use a trivial system in our illustration.  We do so to make
the description more pedagogical. However, it is important to think
about how the method generalizes. This part of the notes should
be used to complement Ref \cite{cpmc0K}, in which more formal and
rigorous discussions can be found. In addition, details of the
algorithm that we will not be able to fully cover here can be found in
those papers.

\subsection{Background}

Here we introduce some background materials and illustrate them with
a simple example. Some of these have been discussed by A.~Muramatsu in
his lecture\cite{Muramatsu}.

\subsubsection{The Hubbard Hamiltonian}

The one-band Hubbard model is a simple paradigm of a system of
interacting electrons. Its Hamiltonian is given by
\begin{equation}
  H = K+V =-t\sum_{\langle ij \rangle \sigma} (c_{i \sigma}^\dagger
    c_{j\sigma} + c_{j \sigma}^\dagger c_{i\sigma}) + U \sum_i n_{i
    \uparrow} n_{i \downarrow},
\label{eq:H}
\end{equation}
where $t$ is the hopping matrix element, and $c_{i \sigma}^\dagger$
and $c_{i \sigma}$ are electron creation and destruction operators,
respectively, of spin $\sigma$ on site $i$. The notation
$\langle\;\rangle$ indicates near-neighbors.  The on-site Coulomb
repulsion is $U>0$, and $n_{i\sigma}=c_{i \sigma}^\dagger c_{i\sigma}$
is the electron number operator.  We will denote the number of lattice
sites by $N$, and the linear dimension by $L$. The numbers of
electrons with spin $\sigma=\uparrow,\,\downarrow$, which we denote by
$N_\uparrow$ and $N_\downarrow$, will be fixed in each
calculation. That is, we will work in the canonical ensemble.  We will
set $t=1$ and impose periodic boundary conditions.

\subsubsection{The Hubbard-Stratonovich transformation}

We now introduce Hirsch's discrete Hubbard-Stratonovich (HS)
transformation\cite{Hirsch}. For any positive $\Delta\tau $, which we will 
choose to be small, the following identity holds:
\begin{equation}
   e^{-\Delta\tau U n_{i \uparrow} n_{i \downarrow}} =
   e^{-\Delta\tau U (n_{i\uparrow}+n_{i \downarrow})/2} \sum_{x_i=\pm 1}
     p(x_i) e^{\gamma x_i (n_{i \uparrow}-n_{i \downarrow})},
\label{eq:HSdiscrete}
\end{equation}
where $\gamma$ is
determined by $\cosh(\gamma) = {\rm exp}(\Delta\tau U/2)$. For reasons
that will become clear in the next section, we have inserted a
function $p(x_i)$ in place of the constant factor $1/2$ and will
interpret $p(x_i)$ as a (discrete) probability density function of
$x_i$, with $x_i=\pm1$.

The essence of the HS transformation is the conversion of an
interacting system into many {\it non-interacting} systems living in
fluctuating external auxiliary-fields, and the summation over all such
auxiliary-field configurations recovers the many-body interactions.
More specifically, the goal is to write the many-body operator
$e^{-\Delta\tau H}$ in terms of one-electron operators.  In Equation
(\ref{eq:HSdiscrete}), the exponent on the left, which comes from the
interaction term $V$, is quadratic in $n$, indicating the interaction
of {\em two\/} electrons. The exponents on the right, on the other
hand, are linear in $n$, indicating two non-interacting electrons in a
(common) external field characterized by $x_i$.

The HS transformation above is one special case for the Hubbard
interaction. Since our purpose is to illustrate the CPMC algorithm
with an example, we will not dwell on more general forms of HS
transformations\cite{HS} here, except to state that they exist for various
other forms of interactions. In general, after an HS transformation
is applied, the propagator $e^{-\Delta\tau H}$ can be written as
\begin{equation}
      e^{-\Delta\tau H}= \int d\vec x\,  p(\vec x) B(\vec x),
\label{eq:HS}
\end{equation}
where $p(\vec x)$ is a probability density function and $B(\vec x)$
is a one-electron operator whose matrix elements are functions of
the many-dimensional auxiliary-fields $\vec x$.

\subsubsection{A specific system to illustrate things}

We now look in some details into the ground-state of a simple
$2\times2$ system of the Hubbard model to help explain the
``language'' we will use when describing CPMC later. We will consider
$N_\uparrow=2$ and $N_\downarrow=1$. We label the four sites 1 thru 4
such that site 1 is near-neighbors with sites 2 and 3.

First let us examine the trivial case of free electrons, i.e., we set
$U=0$. We can write down the {\em 1-electron\/} Hamiltonian matrix,
which is of dimension $4\times 4$:
\begin{displaymath}
H = \left(\begin{array}{cccc} 0 & -1 & -1 & 0\\
                    -1 & 0 & 0 & -1  \\
                    -1 & 0 & 0 & -1  \\
                    0 & -1 & -1 & 0\end{array}\right).
\end{displaymath}
The eigenstates of $H$ can be obtained by direct diagonalization.
With these eigenstates, we immediately obtain the ground-state
wave function $|\psi_0\rangle$ of the 3-electron system from 
the Pauli exclusion principle:
\begin{displaymath}
|\psi_0\rangle = \left(\begin{array}{cc} 0.5 & -0.58 \\
                    0.5 & 0.40 \\
                    0.5 & -0.40 \\
                    0.5 & 0.58\end{array}\right)
\otimes
\left(\begin{array}{c} 0.5\\
                       0.5\\
                       0.5\\
                       0.5\end{array}\right),
\end{displaymath}
where the first matrix contains two single-particle orbitals (two
columns) for the two $\uparrow$ electrons and the second matrix
contains one single-electron orbital for the one $\downarrow$
electron. Each single-electron orbital is an eigenvector of
$H$. Note the second and third lowest single-electron states are degenerate,
causing the 3-electron system to be open shell and $|\psi_0\rangle$ to
be degenerate. We have simply picked one particular linear combination
for the second $\uparrow$-electron in $|\psi_0\rangle$ above.

An object of the form of $|\psi_0\rangle$ is of course nothing more
than a Slater determinant. For example, the amplitude of the configuration
$|R\rangle=|\downarrow 0 \uparrow\uparrow\rangle$, i.e., two $\uparrow$
electrons on sites 3 and 4 and the one $\downarrow$ electron on site 1,
is given by
\begin{displaymath}
\langle R|\psi_0\rangle =
 {\rm det} \left(\begin{array}{cc} 0.5 & 0.40 \\
                    0.5 & -0.58\end{array}\right)
\cdot
{\rm det} \left(\begin{array}{c}
                       0.5\end{array}\right).
\end{displaymath}
That is, more formally,
\begin{displaymath}
 |R\rangle =
 \left(\begin{array}{cc} 0 & 0 \\
      0 & 0 \\
      1 & 0 \\
      0 & 1\end{array}\right)
\otimes
\left(\begin{array}{c}1\\
                      0\\
                      0\\
                      0\end{array}\right)
\end{displaymath}
and $\langle R|\psi_0\rangle = {\rm det} (R^\dagger \cdot \Psi_0)$, where
$R$ and $\Psi_0$ denote the matrices corresponding to $|R\rangle$
and $|\psi_0\rangle$, respectively.  In general, the overlap of two
Slater determinants
\begin{equation}
  \langle\phi |\phi^\prime\rangle =
{\rm det} (\Phi^\dagger\cdot \Phi^\prime)
\label{eq:overlap}
\end{equation}
is a number.

For this non-interacting system, an alternative (albeit indirect and
indeed circular) way of obtaining $|\psi_0\rangle$ is by the power
method. From the eigenvalues and eigenvectors of $K$, we can easily
construct the matrix for $e^{-\Delta\tau H}$, which has the same
structure as $H$ above (i.e., a $4\times 4$ matrix). Denote this
matrix by $B_K$. With an arbitrarily chosen initial Slater determinant
$|\psi^{(0)}\rangle$ (with non-zero overlap with $|\psi_0\rangle$), we
can then repeatedly apply $e^{-\Delta\tau H}$, which means multiplying
both the $4\times2$ $\uparrow$ matrix and the $4\times 1$ $\downarrow$
matrix by $B_K$. As indicated in Table 1, this leads to $|\psi_0\rangle$
asymptotically, i.e.,
\begin{equation}
  |\psi^{(n+1)}\rangle \propto e^{-\Delta\tau H}|\psi^{(n)}\rangle 
\label{eq:process}
\end{equation}
gives $|\psi_0\rangle$ as $n \to \infty$.

Now suppose we turn on the interaction $U$. The first approach of
directly diagonalizing $H$ is the method of exact diagonalization,
which scales exponentially. The power method of
Eq.~(\ref{eq:process}), on the other hand, can still apply if we can
write $e^{-\Delta\tau H}$ in some one-electron form.  The HS
transformation does just that. Assuming $\Delta\tau$ is small and
applying the Trotter break-up, we have
\begin{eqnarray*}
 e^{-\Delta\tau H} = \sum_{\vec x} p(\vec x) & \ 
 & \left(\begin{array}{cccc} e^{\gamma x_1} & 0 & 0 & 0 \\
      0 & e^{\gamma x_2} & 0 & 0 \\
      0 & 0 & e^{\gamma x_3} & 0 \\
      0 & 0 & 0 & e^{\gamma x_4}\end{array}\right)\cdot B_K \\
 & \otimes
 & \left(\begin{array}{cccc} e^{-\gamma x_1} & 0 & 0 & 0 \\
      0 & e^{-\gamma x_2} & 0 & 0 \\
      0 & 0 & e^{-\gamma x_3} & 0 \\
      0 & 0 & 0 & e^{-\gamma x_4}\end{array}\right)\cdot B_K,
\end{eqnarray*}
where $\vec x= \{x_1,x_2,x_3,x_4\}$. This is just Eq.~(\ref{eq:HS}).
Note that $B(\vec x)$ has an $\uparrow$ and a $\downarrow$ component,
each of which is a $4\times 4$ matrix. Applying each $B(\vec x)$ to a
Slater determinant means precisely the same as in the non-interacting
case (with $B_K\otimes B_K$). In other words, $B(\vec x)$ operating on any
Slater determinant $|\phi\rangle$ simply involves matrix multiplications
for the $\uparrow$ and $\downarrow$ components separately, leading to
another Slater determinant $|\phi^\prime\rangle$:
\begin{equation}
   |\phi^\prime\rangle = B(\vec x)|\phi\rangle.
\label{eq:Btimesphi}
\end{equation}

We now have all the basics to understand the projector QMC algorithm
described in Table 1, and to move on to discuss CPMC.

\subsection{Random walk in Slater determinant space}

The first component of the CPMC algorithm is the reformulation of the
projection process as branching, open-ended random walks (a la GFMC) in
Slater determinant space.

We start from the projection process in Eq.~(\ref{eq:process}).
Using (\ref{eq:HS}), we write it
\begin{equation}
   |\psi^{(n+1)}\rangle = \int d\vec x\, p(\vec x) B(\vec x)|\psi^{(n)}\rangle.
\label{eq:cpmc}
\end{equation}
In the Monte Carlo realization of this iteration, we represent the
wave function at each stage by a finite ensemble of Slater
determinants, i.e.,
\begin{equation}
|\psi^{(n)}\rangle \propto \sum_k |\phi^{(n)}_k\rangle,
\label{eq:wf}
\end{equation}
where $k$ labels the Slater determinants and an overall normalization
factor of the wave function has been omitted. Analogous to GFMC, the
Slater determinants will be referred to as {\it random
walkers\/}. We note that, contrary to the walkers in GFMC, the walkers here
are non-orthogonal. Further, their characterization is not as simple
as that of a point, because various
combinations of single-particle orbitals can lead to the same Slater
determinant.

The iteration in (\ref{eq:cpmc})
is achieved stochastically by Monte Carlo (MC) sampling of
$\vec x$:
\begin{equation}
|\phi^{(n+1)}_k\rangle
\leftarrow
\int d\vec x\,p(\vec x) B(\vec x) |\phi_k^{(n)}\rangle;
\label{eq:cpmcsampling}
\end{equation}
that is, for each random walker we choose an
auxiliary-field configuration $\vec x$ from the probability density
function $p(\vec x)$ and  propagate the walker to a new one via
$|\phi^{(n+1)}_k\rangle=B(\vec x)|\phi^{(n)}_k\rangle$. We repeat
this procedure for {\it all\/} walkers in the population.  These
operations accomplish one step of the random walk. The new population
represents $|\psi^{(n+1)}\rangle$ in the sense of
(\ref{eq:wf}), i.e., $|\psi^{(n+1)}\rangle \propto \sum_k
|\phi_k^{(n+1)}\rangle$.
These steps are iterated indefinitely. After an equilibration phase,
all walkers thereon are MC samples of the ground-state wave function
$|\psi_0\rangle$ and ground-state properties can be computed.

The algorithm we have described so far, while correct, is not
efficient, because the sampling of $\vec x$ is completely random with
no regard to the potential contribution to the ground-state wave
function.  In order to improve the efficiency of (\ref{eq:cpmc}) and
make it a practical algorithm, an importance sampling\cite{GFMC} scheme is
required.  To further motivate the need for importance sampling and
develop some insights on how to proceed, we consider the so-called
mixed estimate of the ground-state energy.  As in GFMC, this is given
by $E_0\equiv \langle \psi_T |H|\psi_0\rangle/\langle \psi_T |
\psi_0\rangle$.  Estimating $E_0$ requires estimating the denominator
by $\sum_\phi \langle\psi_T | \phi\rangle$, in which $|\phi\rangle$
denotes random walkers after equilibration. Hence, the term
$\langle\psi_T | \phi\rangle$ plays the role of the Boltzmann
distribution in statistical physics, while the denominator resembles
the partition function. Our current way of sampling corresponds to
sampling completely randomly configurations from the ensemble, which
would clearly lead to large statistical fluctuations in the evaluation of the
partition function. To improve the
situation, it is natural to draw the analogy with the standard
practice of Monte Carlo in statistical physics and try to sample
$|\phi\rangle$ according to $\langle\psi_T | \phi\rangle$. This way,
each walker would ideally contribute equally to the estimate, and
$E_0$ would be given by the average of the quantity $\langle
\psi_T|H|\phi\rangle /\langle \psi_T|\phi\rangle$ with respect to the
sampled walkers.  For a good $|\psi_T\rangle$, fluctuations in this
quantity are small and we expect the statistical error to be much
reduced.

With importance sampling, we iterate a modified equation with
a modified wave function, without changing the underlying eigenvalue
problem of (\ref{eq:cpmc}). Specifically, for each Slater determinant
$|\phi\rangle$, we define an importance function
\begin{equation}
O_T(\phi)\equiv \langle \psi_T|\phi\rangle,
\label{eq:O_T}
\end{equation}
which estimates its overlap with the ground-state wave function. We can
then rewrite equation (\ref{eq:cpmc}) as
\begin{equation}
   |\tilde \psi^{(n+1)}\rangle = \int d\vec x \tilde p(\vec x) B(\vec x)|
      \tilde \psi^{(n)}\rangle,
\label{eq:impcpmc}
\end{equation}
where the modified ``probability density function'' is
\begin{equation}
\tilde p(\vec x)={O_T(\phi^{(n+1)})
            \over O_T(\phi^{(n)})} p(\vec x).
\label{eq:impcpmc_p}
\end{equation}
As expected, $\tilde p(\vec x)$ is a function of both the current and
future positions in Slater-determinant space. Further, it modifies
$p(\vec x)$ such that the probability is increased when $\vec x$ 
leads to a determinant with larger overlap and is decreased otherwise.
It is trivially verified that equations
 (\ref{eq:cpmc}) and (\ref{eq:impcpmc}) are identical.

In the random walk, the ensemble of walkers
$\{\,|\phi^{(n)}_k\rangle\,\}$ now
represents the modified wave
function: $|\tilde \psi^{(n)}\rangle  \propto \sum_k |\phi_k^{(n)}\rangle$.
The true wave function is then given formally by
\begin{equation}
|\psi^{(n)}\rangle  \propto \sum_k |\phi_k^{(n)}\rangle/O_T(\phi_k^{(n)}),
\end{equation}
although in actual measurements it is  $|\tilde \psi^{(n)}\rangle$ that is
needed and division by $O_T$ does not appear.
The iterative
relation for each walker is again given by (\ref{eq:cpmcsampling}),
but with $p(\vec x)$ replaced by $\tilde p(\vec x)$.  The latter
is in general not a normalized probability density function, and we
denote the normalization constant for walker
$k$ by $N(\phi_k^{(n)})$ and rewrite (\ref{eq:cpmcsampling}) as
\begin{equation}
|\phi^{(n+1)}_k\rangle
\leftarrow
N(\phi_k^{(n)}) \int d\vec x\,\frac{\tilde p(\vec x)}{N(\phi_k^{(n)})}
B(\vec x) |\phi_k^{(n)}\rangle.
\label{eq:cpmcimpsamp}
\end{equation}
This iteration now forms the basis of the CPMC
algorithm. As in the DMC method, it is convenient to associate a weight $w_k^{(\
n)}$ with
each walker, which can be initialized to unity. 
With the introduction of
the weight, $|\tilde\psi^{(n)}\rangle \propto \sum_k w_k^{(n)}
|\phi_k^{(n)}\rangle$.
For each walker
$|\phi_k^{(n)}\rangle$, one
step of the algorithm is then as follows: 

\begin{enumerate}

\item sample a $\vec x$ from
the probability density function $\tilde p(\vec x)/N(\phi_k^{(n)})$,

\item propagate the walker by $B(\vec x)$ to generate
a new walker, 

\item compute a weight $w_k^{(n+1)}=
w_k^{(n)}N(\phi_k^{(n)})$ for the new walker. 

\end{enumerate}

To better see the effect of importance sampling, we observe that if
$|\psi_T\rangle=|\psi_0\rangle$, the normalization
$\int \tilde p(\vec x)d\vec x$
is constant. Therefore the weights of walkers remain
a constant and the random walk has no fluctuation. Furthermore, we
refer again to the estimator for $E_0$. With
importance sampling, the denominator becomes the
sum of weights $w$, while the numerator is $\sum_{\phi} \langle
\psi_T|H|\phi\rangle
w_\phi/\langle\psi_T|\phi\rangle$, where again $|\phi\rangle$ denotes
walkers after equilibration. As
$|\psi_T\rangle$
approaches $|\psi_0\rangle$, all walkers contribute with equal weights to the
estimator and the variance approaches zero.

We stress again that the importance sampling transformation does
not alter the underlying iterative equation in (\ref{eq:cpmc}). The
results of the calculations are not affected; different choices of the
importance function only affects the efficiency of the algorithm
and thus the statistical error.

\subsection{The sign problem and the constrained path approximation}

The second component of the CPMC algorithm is the constrained path
approximation to eliminate the sign decay.

As a side-product, the reformulation of the projection process
described in 2.2. provides a clear picture of the origin of the sign
problem.  Below we first elaborate on this picture, from which the
constrained path approximation emerges naturally.

The sign problem occurs because of the fundamental symmetry between
the fermion ground-state $|\psi_0\rangle$ and its negative
$-|\psi_0\rangle$\cite{Fahy,inter}.  For any
ensemble of Slater determinants $\{|\phi\rangle\}$ which gives a Monte Carlo
representation of the ground-state wave function, this symmetry
implies that
there exists another ensemble
$\{-|\phi\rangle\}$ which is also a correct representation.  In other
words, the Slater determinant space can be divided into two degenerate
halves ($+$ and $-$) whose bounding surface ${\cal N}$ is defined by
$\langle\psi_0|\phi\rangle=0$. This surface 
is in general {\it unknown\/}.

In some special cases, such as the particle-hole symmetric,
half-filled one-band Hubbard model, symmetry prohibits any crossing of
${\cal N}$ in the random walk. The calculation is then free of the
sign problem\cite{Hirsch}.
In more general cases, however, walkers do cross ${\cal N}$ in their
propagation by $e^{-\Delta\tau H}$. The sign problem then invariably
occurs.

\begin{figure}
\epsfysize=3.2in
\vskip-1.4truein
\centerline{\epsfbox{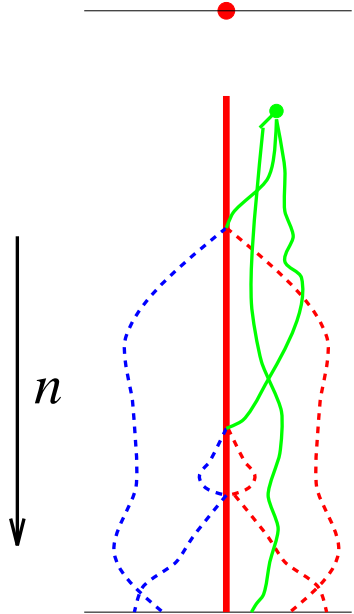}}
\vspace{1.6in}
\caption{Schematic illustration of the sign
problem.  The top line represents Slater determinant space; the
dot represents the ``node'' ${\cal N}$, where a
determinant is orthogonal to the ground state $|\psi_0\rangle$.  As
the projection continues (increasing $n$), Slater determinants undergo
random walks, tracing out ``paths'' as shown. When a walker reaches
${\cal N}$, its future paths will {\em collectively\/} cancel in their
contribution to $|\psi_0\rangle$, indicated by the symmetric distribution
of dashed paths about the nodal line. The Monte Carlo sampling, with no
knowledge of this cancellation, continues to sample such paths
randomly. The relative number of paths with real contributions (solid
paths) decreases {\em exponentially\/} as $n$ increases. }
\label{fig_sign}
\end{figure}

In Fig.~\ref{fig_sign}, we illustrate the space of Slater
determinants by a one-dimensional (horizontal) line. The node ${\cal
N}$ is the (red) dot in the middle.  Imaginary time (or $n$) is in the
vertical direction and increases as the arrow suggests. That is, as
the walker moves in the horizontal line, we stretch out continuously
``snapshots'' of its position along the vertical direction.  Now we
follow an initial Slater determinant. With no loss of generality, we
assume it has a positive overlap with $|\psi_0\rangle$.  At time $n=0$ it
is indicated by the (green) dot on the right. As the random walk
evolves, the walker can reach the node, which is the (red) vertical
line. At the instant it lands on ${\cal N}$, the walker will make no
further contribution to the representation of the ground state, since
\begin{equation}
\langle \psi_0|\phi\rangle = 0 \ \Rightarrow \
\langle \psi_0|e^{-\tau H}|\phi\rangle
= 0 \ {\rm for \ any} \ \tau.
\label{eq:node}
\end{equation}
Paths that result from such a walker have equal probability of being
in either half of the Slater determinant space. A few of these
possible paths are shown by dashed lines.  Computed analytically, they
would cancel and not make any contribution in the ground-state wave
function, as indicated by their symmetric placement with respect to 
the node line. But since the random walk has no knowledge of ${\cal N}$,
these paths continue to be sampled (randomly) in the random walk and
become Monte Carlo noise. Only paths that are completely confined to
the right-hand side, as shown by the solid (green) line, will lead to
contributions to the ground state, but the relative number of such
confined paths decreases exponentially with $n$.  Asymptotically in 
$n$, the Monte Carlo representation of the ground-state wave
function consists of an {\it equal\/} mixture of the $+$ and $-$
walkers, regardless of where the random walks originated. The Monte
Carlo signal is therefore lost. The decay of the signal-to-noise
ratio, i.e. the decay of the average sign of
$\langle\psi_T|\phi\rangle$, occurs at an exponential rate with
imaginary time.

To eliminate the decay of the signal-to-noise ratio, we impose the
con-strained-path approximation. Fahy and Hamann first used\cite{Fahy}
such a constraint in the framework of the standard AFQMC
method. However, the non-local nature of such a constraint proved
difficult to implement efficiently in the ``path-integral-like''
projector QMC scheme.  Here, with the random walk formalism, the
constraint only needs to be imposed one time-step at a time and is
extremely simple to implement.  It requires that each random walker at
each step have a positive overlap with the trial wave function
$|\psi_T\rangle$:
\begin{equation}
\langle \psi_T|\phi_k^{(n)}\rangle > 0.
\label{eq:constraint}
\end{equation}
This yields an approximate solution to the ground-state
wave function, $|\psi^c_0\rangle =\sum_\phi |\phi\rangle$, in which
all Slater determinants $|\phi\rangle$ satisfy
(\ref{eq:constraint}).

 From (\ref{eq:node}), it follows that the constrained path
approximation becomes {\it exact\/} for an exact trial wave function
$|\psi_T\rangle=|\psi_0\rangle$. The overall
normalization of walkers remains a constant on the average, since the
loss of walkers at ${\cal N}$ is compensated by the branching of
walkers elsewhere; that is, the eigenvalue problem with ${\cal N}$ as
boundary has a stable solution.

To implement the constrained-path approximation in the random walk, we
redefine the importance function by (\ref{eq:O_T}):
\begin{equation}
O_T(\phi)\equiv {\rm max}\{ \langle \psi_T| \phi\rangle,
0\}.
\label{eq:imp_cp}
\end{equation}
This prevents walkers from crossing the trial nodal surface ${\cal N}$
and entering the ``$-$'' half-space as defined by
$|\psi_T\rangle$.
In the limit $\Delta\tau\rightarrow 0$, (\ref{eq:imp_cp}) ensures that
the walker distribution vanishes smoothly at ${\cal N}$ and the
constrained-path approximation is properly imposed. With a finite
$\Delta \tau$, however, $\tilde p(\vec x)$ has a discontinuity at ${\cal
N}$ and the distribution does not vanish. We have found this
effect to be very small for reasonably small imaginary-time steps
$\Delta\tau$ . Nonetheless, we correct for it by modifying
$\tilde p(\vec x)$ near ${\cal N}$ so that it approaches zero smoothly
at ${\cal N}$.

\subsection{Additional algorithmic issues}

\subsubsection{Computing expectation values}

Given the ground-state wave function $|\psi_0^{\rm c}\rangle$ from
CPMC, where the superscript ``c' indicates that the wave function is
obtained under the constrained-path approximation, we can compute the
ground-state energy with the mixed estimate\cite{GFMC}:
\begin{equation}
E_{\rm CPMC}^{\rm mixed} = {\langle \psi_T|H|\psi_0^{\rm c}\rangle
\over \langle \psi_T|\psi_0^{\rm c}\rangle}.
\label{eq:mixed}
\end{equation}
This can be implemented in a way similar to GFMC.  However, contrary
to fixed-node in GFMC, the ground-state energy computed from the mixed
estimate is {\em not\/} an upper bound\cite{upper}. (The argument
given in Ref.~\cite{cpmc0K} on variational properties is incorrect.)
The origin for the disappearance of the upper bound property in the
mixed-estimate is the non-orthogonal nature of the Slater determinant
space.  In contrast with configuration space in fixed-node GFMC, where
points on the node have no overlap with either $|\psi_T\rangle$ or the
fixed-node solution, a point in the Slater determinant space which has
no overlap with $|\psi_T\rangle$ is not necessarily orthogonal with
$|\psi_0^{\rm c}\rangle$.  Eq.~(\ref{eq:mixed}) is {\em not\/}
equivalent to the estimator $\langle \psi_0^{\rm c}| H|\psi_0^{\rm
c}\rangle/ \langle \psi_0^{\rm c}| \psi_0^{\rm c}\rangle$, which is
variational.

In the Hubbard model, the effect of this appears to be
small\cite{upper}. In Table I, we show several cases we could find
where the computed ground-state energies by CPMC fell below the exact
results. The system is $4\times 4$, with $5 \uparrow$ and $5
\downarrow$ electrons, which corresponds to a special case of a
closed-shell system in the non-interacting limit.  The free-electron
wave function was used as the trial $|\psi_T\rangle$.
\begin{table}[htb]
\begin{center}
\caption{Examples showing the lack of upper bound property in the {\em mixed-estimate\/} 
of CPMC. The system is $4\times 4$, with $5 \uparrow$ and $5
\downarrow$ electrons. Calculations were done for three values 
of the interaction strength $U$, $4$, $8$, and $20$.
The CPMC results are obtained from the
mixed estimate; statistical errors are in the 
last digit and are indicated in parentheses. These are compared with 
results from exact 
diagonalization.}
\begin{tabular}{clll}
\hline
$U$ & 4 & 8 & 20 \\
\hline
CPMC  & -19.582(5) & -17.519(2) & -15.460(3)\\
exact & -19.580    & -17.510    & -15.452\\
\hline
\end{tabular}
\end{center}
\end{table}
As we see, the amount by which the mixed estimates fall below the
exact results is no more than $0.1\%$ across a wide range of
interactions.  For $U=8$, an ``almost'' free-electron wave function
was used as $|\psi_T\rangle$ in Ref.~\cite{cpmc0K}, namely an
unrestricted Hartree-Fock wave function generated with a small
$U$. That $|\psi_T\rangle$ led to a ground-state energy {\em above\/}
the true value\cite{cpmc0K}.

Several possibilities exist to
correct for the non-variational
nature of the mixed estimate, at least 
in principle, and make it an upper bound\cite{upper}.  
For example, we could compute
$\langle \psi_0^{\rm c}| H|\psi_0^{\rm c}\rangle/ \langle \psi_0^{\rm
c}| \psi_0^{\rm c}\rangle$ directly with a method similar to forward
walking\cite{fwd,nigh} in GFMC. That is, we can create a separate walk for the left-hand wave
function $\langle \psi_0^{\rm c}|$, which propagates from $\langle
\psi_T|$ with the constraint properly implemented, and then compute
the overlap with populations in the regular (right-hand) walk. It
would clearly be inefficient to {\em randomly\/} sample the left-hand
walks, whose overlap with the regular walks would fluctuate
greatly. We have devised an ``importance sampling'' scheme that both
imposes the constraint with $| \psi_T \rangle$ and attempts to
incorporate the necessary correlations between the left-hand random
walker $\langle \phi^\prime|$ and its matching right-hand walker. We take an importance
function $O_T^L(\phi^\prime) = | \langle \phi^\prime | \phi_k^{(n)}
\rangle | + \alpha \langle \phi^\prime | \psi_T \rangle$, where
$\alpha$ is a parameter and $|\phi_k^{(n)} \rangle$ is the walker with
which $\langle \phi^\prime|$ will be matched in the regular population
representing the right-hand wave function. While this offered a
significant improvement over naive sampling, we found it still prone
to large fluctuations.  With the variational correction term to
the mixed estimate being very small, the statistical accuracy obtained
by this scheme did not appear adequate to be of practical use for
large systems. We note that, although this way of forward walking is
somewhat similar to the back-propagation technique we discuss below,
they are not equivalent. The latter tends to be much more accurate
statistically. However, as we see below, it does not maintain the
correct sense of direction in imposing the constraint and does not
yield strictly $\langle \psi_0^{\rm c}|$.

The back-propagation (BP) technique enables the computation of
the expectation value of an operator ${\cal O}$ that does not commute with $H$. 
As we have mentioned, the
essence of BP also comes from the forward walking 
technique in the GFMC method:
\begin{equation} \langle{\cal O}\rangle_{\rm BP} = \lim_{\tau \rightarrow
 \infty}
\frac{\langle \psi_T \exp ( - \tau H_c)| {\cal O}| \psi_0^c \rangle}
     {\langle \psi_T \exp ( - \tau H_c)| \psi_0^c \rangle}.
\label{eq:bpest}
\end{equation}
A subtle distinction, however, exists between back-propagation and
forward walking.  In back-propagation, $\langle \psi_T \exp ( - \tau
H_c ) | = \langle \psi_T| \exp ( -\tau H_c )$ is restricted to
``constrained'' paths, i.e., those paths that do not violate the
constraint in the {\it original forward direction} $\exp ( -
\Delta\tau H_c )|\psi_0^c \rangle$.  In fixed-node GFMC a path in
configuration space has no sense of direction with respect to the
nodal surface. In the CPMC method, however, there is a sense of
direction: a set of determinants along the path of a random walk which
does not violate the constraint at any step when going from right to
left may indeed violate it any even number of times when going from
left to right.  Because of this sense of direction, expression
(\ref{eq:bpest}) may not yield the true expectation value $\langle
\psi_0^{\rm c}| {\cal O}|\psi_0^{\rm c}\rangle/ \langle \psi_0^{\rm
c}| \psi_0^{\rm c}\rangle$. However, since $|\psi_0^c\rangle $ is
itself approximate, this issue is not crucial. 
$\langle{\cal O}\rangle_{\rm BP}$ does have the correct limiting
behavior, remaining exact for an exact trial wave function.

It is relatively simple to implement the back-propagation scheme on
top of a regular CPMC calculation. The key is that once a path has been 
sampled, the single-particle propagator can be easily constructed.
Propagating a Slater determinant with it, either forward or backward,
simply means matrix multiplications.
We choose an iteration $n$
and store the entire population
$\{\,|\phi^{(n)}_k\rangle\,\}$.  As the random walk proceeds, we keep
track of the following two items for each new walker: (1) the sampled
auxiliary-field variables that led to the new
walker from its parent walker and (2) an integer that labels the
parent. After an additional $m$ iterations, we carry out the
back-propagation: For each walker $l$ in the $(n+m)^{\rm
th}$ (current) population, we initiate a determinant $\langle \psi_T|$ and act
on it with the corresponding propagators, but taken in reverse
order. The $m$ successive propagators are constructed from the stored
items, with ${\rm exp}(-\Delta\tau K/2)$ inserted where necessary. The
resulting determinants $\langle \bar\phi^{(m)}_l|$ are combined with
its parent from iteration $n$ to compute $\langle{\cal O}\rangle_{\rm
BP}$, in a way similar to the mixed estimator (\ref{eq:mixed}). The
weights are given correctly by $w_l^{(n+m)}$, due to importance
sampling in the regular walk. Starting from another iteration $n'$,
this process can be repeated and the
results accumulated.

We note again that, because the random walkers in CPMC are full Slater
determinants, the algorithm lends itself well to the calculation of
expectation values. The non-orthogonal nature of the determinant space
ensures that overlaps between two walkers can be computed. This is to
be contrasted with `'continuum'' methods in Table 1, where
off-diagonal expectations are in general difficult to compute, because
it is difficult to sample two group of random walkers whose overlap 
is well behaved statistically.

\subsubsection{population control and bias correction}

In CPMC walkers carry weights or, equivalently, they branch.  The
procedure we use to control the population is similar to that used in
many GFMC calculations.  First, a branching (or birth/death) scheme is
applied, in which walkers with large weights are replicated and ones
with small weights are eliminated by probabilities defined by the
weights. There exist various ways to do
this\cite{Umrigar,nigh,Hetherington}, with the guideline being that
the process should not change the distribution statistically. In 
general, how this step is done only affects the efficiency of the 
algorithm, but does not introduce any bias.

Branching allows the total number of walkers to fluctuate and possibly
become too large or too small. Thus as a second step, the population
size is adjusted, if necessary, by rescaling the weights with an
overall factor. Re-adjusting the population size, i.e., changing the
{\em overall\/} normalization of the population, does introduce a
bias\cite{Umrigar,nigh}.  We correct for this bias by carrying the $m$
most recent overall rescaling factors and including them in the
estimators when computing expectation values.  In the calculation we
keep a stack which stores the $m$ latest factors, $f^{(j)}$ ($j=1,m$).
Suppose that at the current step the total number of walkers exceeds
the pre-set upper bound. We modify the weight of each walker by a
constant factor $f<1$ which reduces the population size to near the
expected number. We then replace the oldest $f^{(j)}$ in the stack by
$f$. Whenever we compute expectation values, we multiply the weight of
each walker by $1/\prod_j f^{(j)}$. In our calculations on the Hubbard
model, $m$ is typically between $5$ and $10$.  As we include more such
factors, i.e., increasing $m$, the bias is reduced, but the
statistical error increases. On the other hand, as we reduce $m$, the
statistical error becomes smaller, but the bias increases. The choice
of $m$ is thus a compromise between these two.  Another common
approach to eliminate or reduce bias, which is in general less
efficient, is to do several calculations with different (average)
population sizes and extrapolate to the infinite population limit.

Clearly, the schemes we have described to correct for bias have some
arbitrariness and further improvements are possible. But this should
not be a major factor in the calculation. In cases where the bias is
{\em substantially\/} larger than can be handled by correction schemes
in this spirit, it is likely more productive to attempt to
improve the importance sampling and the algorithm, rather than details
of the bias correction scheme.

\subsubsection{re-orthogonalization a Slater determinant}

Repeated multiplications of $B(\vec x)$ to a Slater determinant,
either in the regular walk or in BP, lead to a numerical instability,
such that round-off errors dominate and $|\phi_k^{(n)}\rangle$
represents an unfaithful propagation of $|\phi_k^{(0)}\rangle$.  This
instability is well-known in the AFQMC method and is
controlled\cite{white,LGbook} by a numerical stabilization technique
that requires the periodic re-orthonormalization of the single
particle orbitals in $|\phi_k^{(n)}\rangle$. 

We use the {\it modified} Gram-Schmidt procedure\cite{LGbook,white} to
stabilize the Slater determinants. The procedure is further simplified
because of the random walk formalism in CPMC.  For each walker
$|\phi\rangle$, we factor the matrix $\Phi$ as $\Phi = QR$, where $Q$
is a matrix whose columns are a set of orthonormal vectors
representing the re-orthogonalized single-particle orbitals. $R$ is a
triangular matrix. Note that $Q$ contains all the information about
the walker $|\phi\rangle$. $R$, on the other hand, only contributes to
the overlap of $|\phi\rangle$. Since the overlap is already
represented by the walker weight, $R$ does not need to be carried
along and can be discarded.

It is interesting to note that this instability originates from the
tendency for each walker (Slater determinant) to collapse to a bosonic
ground state. In GFMC this actually happens and is the first cause for
the sign problem. Here the instability can be eliminated by the
re-orthonormalization procedure because a walker is explicitly a
Slater determinant, not a single point. The procedure to stabilize the
Slater determinants is like analytically canceling walkers in
GFMC\cite{inter}, which requires a large density of walkers and does
not scale well with system size. CPMC, like AFQMC, automatically
imposes the antisymmetry. This would suggest that the sign problem is
reduced in this formalism compared to GFMC. One trivial observation
which is consistent with this is the case of a non-interacting system,
where CPMC does not suffer from the sign problem and has zero-variance,
while GFMC does have a sign problem and still requires the fixed-node
approximation.

\subsection{Illustrative benchmark results}

Here we show a few benchmark results to illustrate characteristics of
the CPMC algorithm. These results are for the two-dimensional
Hubbard model. Ref.~\cite{cpmc0K} contains more results and a more
extensive account of the benchmark calculations that have been
performed.

\begin{figure}
\vspace{1.8in}
\epsfxsize=5.3in
\epsfysize=4.in
\centerline{\epsfbox{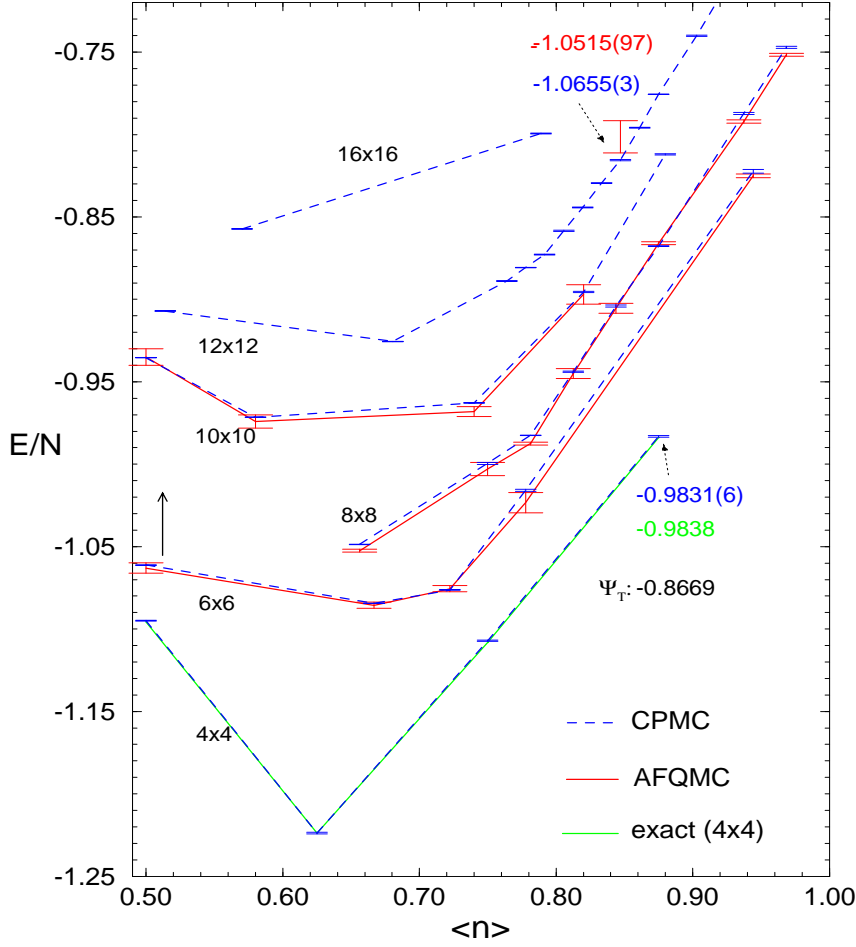}}
\vspace{-0.6in}
\caption{
Computed energies per site vs.\ electron fillings from CPMC, compared
with available AFQMC and exact diagonalization data. 
The lines are to aid the eye. Curves for $L>4$ are shifted up. 
Half-filling is $\langle n\rangle =1$. For $4\times 4$, the solid curve
is from exact diagonalization\protect\cite{diag}. 
At $\langle n\rangle =0.875$ ($7 \uparrow
\ 7 \downarrow$), the CPMC (with error bar in the last digit) and exact 
numbers are shown, together with the variational energy from the 
trial wave function $|\psi_T\rangle$ in CPMC. Numbers are also shown 
for the AFQMC (lone point) and CPMC values for $12\times 12$ with
$61 \uparrow \ 61 \downarrow$. Note that the CPMC error bar is smaller by
a factor of $30$ in this case.}
\label{fig2}
\end{figure}

In Figure \ref{fig2} we show the computed $E_0/N$ at $U=4$ for various
lattice sizes and electron filling $\langle n \rangle=(N_\uparrow
+N_\downarrow)/N$. Also shown is available AFQMC data \cite{Imada}.
We see that the CPMC results are in good agreement with the AFQMC
data. For example, in the case of $8\times 8$ lattice with $25$
$\uparrow$ and $25$ $\downarrow$ electrons, the CPMC result lies
approximately 0.4\% $\pm$ 0.2\% above the AFQMC result. The behaviors
of the statistical errors in these two algorithms are of course
completely different.  In AFQMC the sign problem causes the variance
to increase {\it exponentially} with $N$ and the projection time
$n\Delta\tau$, while the CPMC method is free of any sign decay and
exhibits power law scaling with $N$.  For large systems ($10\times 10$
and beyond), the CPMC error bars are $30$-$50$ times smaller than
those of AFQMC, as indicated for the $12\times 12$ system. 

A more stringent test of the algorithm is the calculation of
correlation functions. In Table 3, we show computed expectation values
for the system of $4\times 4$ with $7\uparrow\,7\downarrow$ electrons,
which is an open-shell case with a severe sign problem. 
The one-body density
matrix is the expectation value of the Green's function
element: $\rho({\bf l})=\langle c_{\bf 0}^\dagger c_{\bf l}\rangle$,
where the vector ${\bf l}=(l_x,l_y)$
denotes the position of a site (vs.~ $i$, which we have used so far and which is 
the site label). The spin density structure factor is
\begin{equation}
S(k_x,k_y) = S({\bf k}) = 1/N\sum_{\bf l} {\rm exp}(i{\bf k}\cdot {\bf l})
\langle {\bf s}_{\bf 0} {\rm s}_{\bf l}\rangle,
\label{eq:struct}
\end{equation}
where ${\bf s}_{\bf l}=n_{{\bf l}\uparrow}-n_{{\bf l}\downarrow}$ is the
spin at site ${\bf l}$.
We show
results from CPMC simulations with two different trial wave
functions. Both are unrestricted Hartree-Fock wave functions, but
$|\psi_{T1}\rangle$ was obtained with a $U$ of $0.1$, while
$|\psi_{T2}\rangle$ with $U=4$. The calculation with
$|\psi_{T1}\rangle$ has much less fluctuation, even though
$|\psi_{T2}\rangle$ has a lower variational energy\footnote{This trend
appears to be rather general: free-electron-like wave functions tend
to be better importance functions than unrestricted Hartree-Fock wave
functions. This likely reflects to what extent translational
invariance is maintained.}.  We see that the two trial wave functions
yield very different variational estimates, but their CPMC results are
consistent and in good agreement with exact results. For
example, in the free-electron-like $|\psi_{T1}\rangle$, the momentum
distribution is a step function, and the ${\bf k}=(1,0)$ state is
completely occupied ($n_k=1$), but even with this trial wave function
the CPMC method still gives the correct occupation of $0.92(1)$.  
The spin structure factor $S(\pi,\pi)$ is the only case where CPMC
does not produce the exact result. However, even with
the free-electron-like $|\psi_{T1}\rangle$, which severely 
underestimates the anti-ferromagnetic correlation, CPMC 
correctly recovers the physics and predicts the presence of
a strong peak of $S({\bf k})$ at $(\pi,\pi)$.

\begin{table}
\begin{center}
\caption{Computed expectation values and correlation functions from
CPMC for a $4\times 4$ lattice with $7\uparrow\,7\downarrow$
electrons and $U=4$,
compared with exact results. Results are shown for two different
trial wave functions $|\psi_{T1}\rangle$ and $|\psi_{T2}\rangle$.
Both are unrestricted Hartree-Fock wave functions,
but $|\psi_{T1}\rangle$ was obtained with a $U$ of $0.1$, while
$|\psi_{T2}\rangle$ with $U=4$.
Exact diagonalization results are from Ref.~\protect\cite{diag}; numbers in
parentheses indicate either the range of values due to the
ground-state degeneracy or uncertainties in extracting numbers from a
graph. Statistical errors on the last digit of the CPMC results are in
parentheses.
 $E_k$ is the kinetic energy, $\rho(l_x,l_y)$ the one-body density
matrix, $S$ the spin density structure factor, and
$n_k$ the momentum distribution.}
\begin{tabular}{r@{\ \ }l r@{}l r@{}l r@{}l r@{}l}
\hline
 & &
\multicolumn{2}{c}{$E_k$} &
\multicolumn{2}{c}{$\rho(2,2)$} &
\multicolumn{2}{c}{$S(\pi,\pi)$} &
\multicolumn{2}{c}{$n_k(\pi/2,0)$} \\ \hline
variational&$|\psi_{T1}\rangle$ &-24.&0 & 1.&654
 & 0.&625 & 1.&0\\
 &$|\psi_{T2}\rangle$ & -21.&88 & -0.&0602 & 4.&39 
 & 0.&941\\ \hline
CPMC & $|\psi_{T1}\rangle$ & -21.&44(2) & -0.&051(1)
 & 2.&90(1) & 0.&92(1)\\
 &$|\psi_{T2}\rangle$ & -21.&39(8) & -0.&049(1) & 2.&92(2)
 & 0.&92(1)\\
\hline
\multicolumn{2}{c}{exact} & -21.&39(1) & -0.&051
 & 2.&16(2) & 0.&93(1)\\
\hline
\end{tabular}
\end{center}
\end{table}

The fact that CPMC shows consistency with different choices of 
$|\psi_T\rangle$ is reassuring. Such consistency checks are 
crucial in real applications where benchmark data is scarce. 
The fact that CPMC appears to be able to accurately calculate
correlation functions in these systems is significant.  Correlation
functions are often much more difficult to compute than the energy. It
is here that the difference in scaling behaviors between CPMC and
AFQMC becomes much more magnified. The AFQMC algorithm tends to break
down for system sizes much smaller than indicated in Fig.~\ref{fig2}.
The fixed-node GFMC also encounters difficulties in computing certain
correlation functions because forward walking does not yield proper
overlaps, as we discussed earlier.

In Table 4, we show a comparison of the computed ground-state energies
from CPMC and from lattice fixed-node GFMC\cite{FNlatt}.  We see that,
in addition to the ability to compute correlation functions
conveniently, CPMC does better even for the energy.  (The particular
system is an easy case for CPMC and the results may not be typical of
the Hubbard model.) At small $U$, CPMC is expected to perform better
because the linear combination of mean-field states provides an
effective description of the system. On the other hand, at very large
$U$, we expect that, at least in the current way of
Hubbard-Stratonovich transformation, the statistical error in CPMC
will be larger than GFMC. More complete studies are necessary to
characterize where the ``cross-over'' occurs.

\begin{table}[htb]
\begin{center}
\caption{Comparison of CPMC with lattice fixed-node (FN) GFMC for 
$4\times 4$, with $5 \uparrow$ and $5
\downarrow$ electrons. 
Shown are computed 
ground-state energies per site.
The CPMC result was obtained with a free-electron $|\psi_T\rangle$,
as was the first number for fixed-node GFMC. The FN GFMC
result ``Gutzwiller'' was obtained with a better $|\psi_T\rangle$
which included a Gutzwiller factor. Statistical errors are in the last digit and 
are indicated in parentheses. Fixed-node results are from 
Ref.~\protect\cite{FNlatt_new}.}
\begin{tabular}{cccc}
\hline
FN GFMC & FN GFMC (Gutzwiller) & CPMC & exact \\
\hline
-1.2186(4) & -1.2201(4) & -1.2239(3) & -1.2238\\
\hline
\end{tabular}
\end{center}
\end{table}

\section{Application: Is the Hubbard model the right one for
high-$T_c$?}

Here we describe an application of the CPMC algorithm to study the
ground state of the one-band Hubbard Hamiltonian of Eq.~(\ref{eq:H}).
In particular, we will focus on the electron pairing correlation in
the $d_{x^2-y^2}$-wave channel. Most of the data in this section is
from Ref.~\cite{pairing}.

The Hubbard model is a simple microscopic description of a
many-electron system which includes both band-structure and electron
interactions.  Since the discovery of high-$T_c$ superconductivity, it
has been widely believed that this model contains essential features
relevant to the properties of the CuO$_2$ planes in the cuprate
superconductors. Viewed as the basis for further progress on
understanding the mechanism for high-$T_c$, the model has 
been the subject of intense theoretical
activity\cite{Dagotto}. 
Various calculations
predict\cite{scalapino95} that the doped model exhibits an attractive
interaction between pairs; the extended $s$- and $d_{x^2-y^2}$-
symmetries of this attraction are consistent with the likely
symmetries of the experimentally measured superconducting gap
\cite{scalapino95}. Yet unobserved, however, is convincing evidence
that the attractive interaction leads to a ground state with
off-diagonal long-range order\cite{Dagotto,scalapino95}. In other 
words, we do not have the complete answer to whether
the ground state of this model has a condensate of electron 
pairs and, if yes, what their nature is. 

Numerical approaches have been a promising tool for answering some of
the questions on the Hubbard model.  The AFQMC method in particular
has been applied extensively and has seen a great deal of
success\cite{scalapino90,Dagotto}.  However, due to the sign problem,
it has typically been limited to relatively small system sizes, high
temperatures, and selected electron fillings. CPMC largely eliminates
these difficulties, although we must bear in mind the
approximate nature of the constrained path approximation.

The pairing correlation function we computed is defined as
\begin{equation}
D_\alpha({\bf l}) = \langle \Delta^\dagger_\alpha({\bf l }) \Delta_\alpha({\bf 0})\rangle,
\label{eq:pairing_def}
\end{equation}
where $\alpha$ indicates the nature of pairing. The pair-field operator
at site ${\bf l}$ is
$\Delta_\alpha ({\bf l}) =
\sum_{\bf \delta} f_\alpha({\bf \delta}) [c_{{\bf l}\uparrow} c_{{\bf l}+{\bf \delta}\,\downarrow}
-c_{{\bf l}\downarrow} c_{{\bf l}+{\bf \delta}\,\uparrow}]$,
 where ${ {\bf \delta}}$
is $(\pm 1,0)$ and $(0,\pm 1)$. For the extended $s$-wave ($s^\star$-), $f_{s^\star} (
{\bf \delta} )=1$. For $d$-wave, $ f_d ({\bf \delta})$ is $1$
when ${ {\bf \delta}}=(\pm 1,0)$ and $-1$ otherwise.

\begin{figure}
\vspace{0.7in}
\epsfxsize=4.in
\epsfysize=3.in
\centerline{\epsfbox{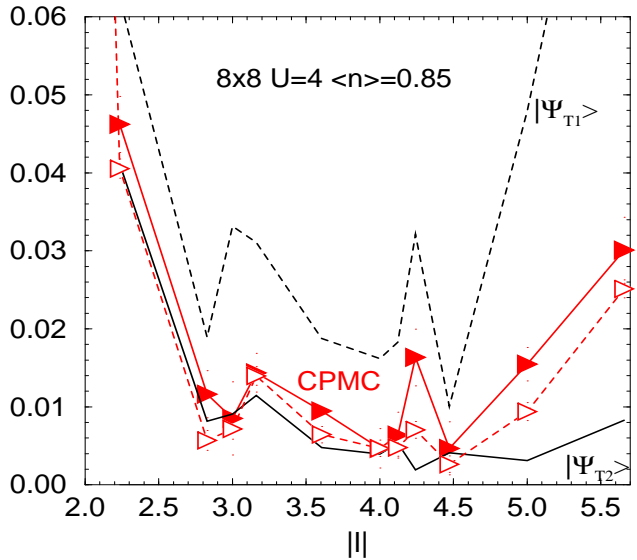}}
\vspace{-0.5in}
\caption{Sensitivity of CPMC results to the choice of trial wave
function $|\Psi_T\rangle$. Shown is the near-neighbor $d$-wave pairing
correlation as a function of pair separation $|{\bf l}|$. The short
distance portion is omitted. This system corresponds to a difficult
open-shell case. The two trial wave functions are both unrestricted
Hartree-Fock wave functions. The first, $|\psi_{T1}\rangle$, was
obtained with a very small interaction strength $U$, resembling a
free-electron wave function, while $|\psi_{T2}\rangle$ was generated
with the true $U$ of $4$. The CPMC pairng correlations, with
$|\psi_{T1}\rangle$ and $|\psi_{T2}\rangle$ as trial wave functions
respectively, are shown (curves with symbols (red)), with the dashed one
from $|\psi_{T1}\rangle$. The mean-field results
that $|\psi_{T1}\rangle$ and $|\psi_{T2}\rangle$ yield are also 
shown (no symbols). We see that the two sets of CPMC results, although
not in exact agreement, over-correct the mean-field results and show
consistency. The statistical error bars on the CPMC
results are indicated by small dots, the maximum being several times the symbol 
size.}
\label{fig3}
\end{figure}

Because the algorithm has an uncontrolled approximation and because
the electron pairing correlations are extremely difficult quantities to
characterize, we have carried out benchmark
calculations\cite{pairing,cpmc0K} and many self-consistency checks.
Due to the limited availability of data for benchmark, the latter
becomes even more important.  In Fig.~\ref{fig3}, we show another such
example, for a fairly difficult case of $8 \times 8$ with 
$27 \uparrow \ 27 \downarrow$ electrons. 

\begin{figure}
\epsfxsize=3.5in
\epsfysize=3.2in
\centerline{\epsfbox{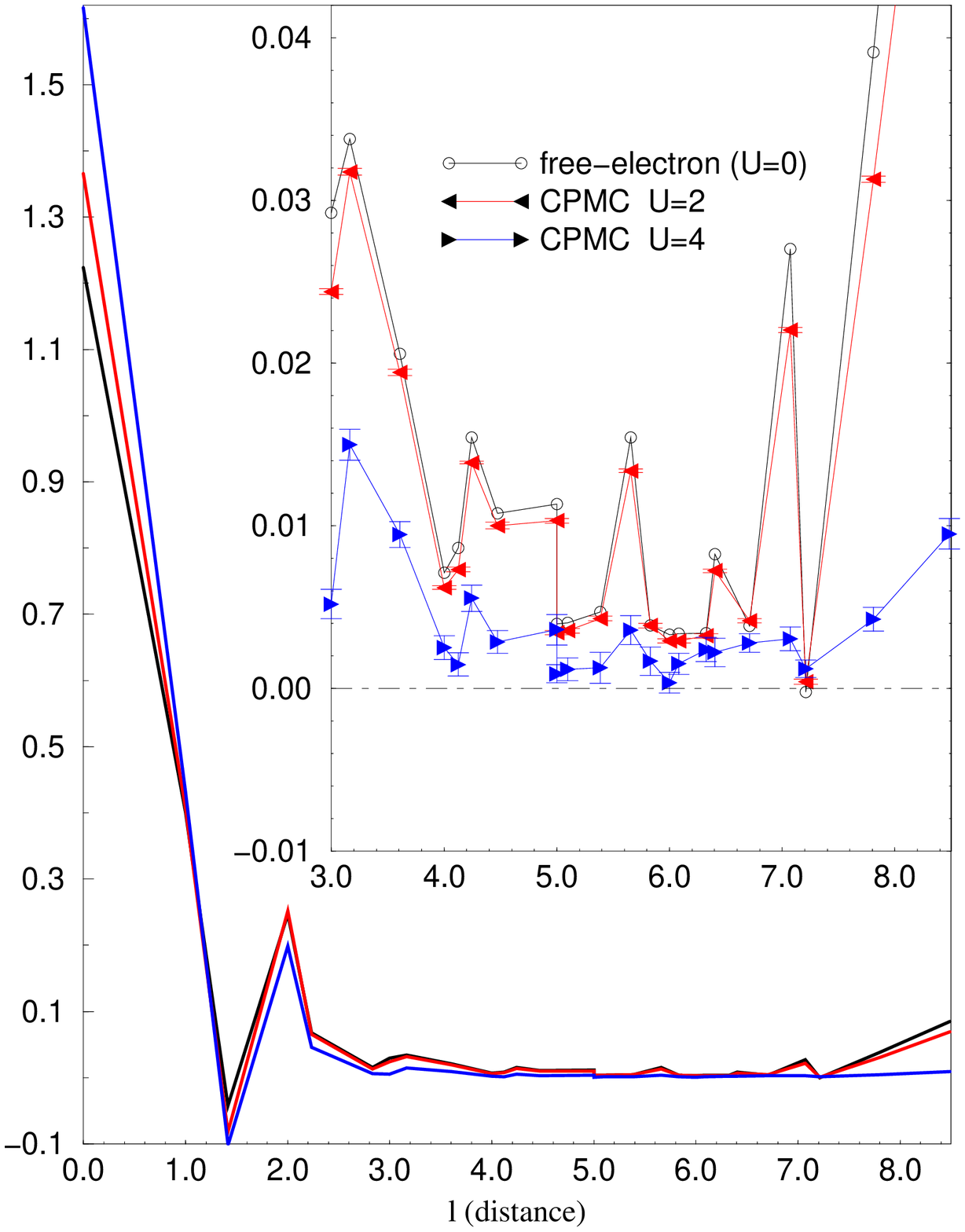}}
\epsfxsize=3.5in
\epsfysize=3.2in
\centerline{\epsfbox{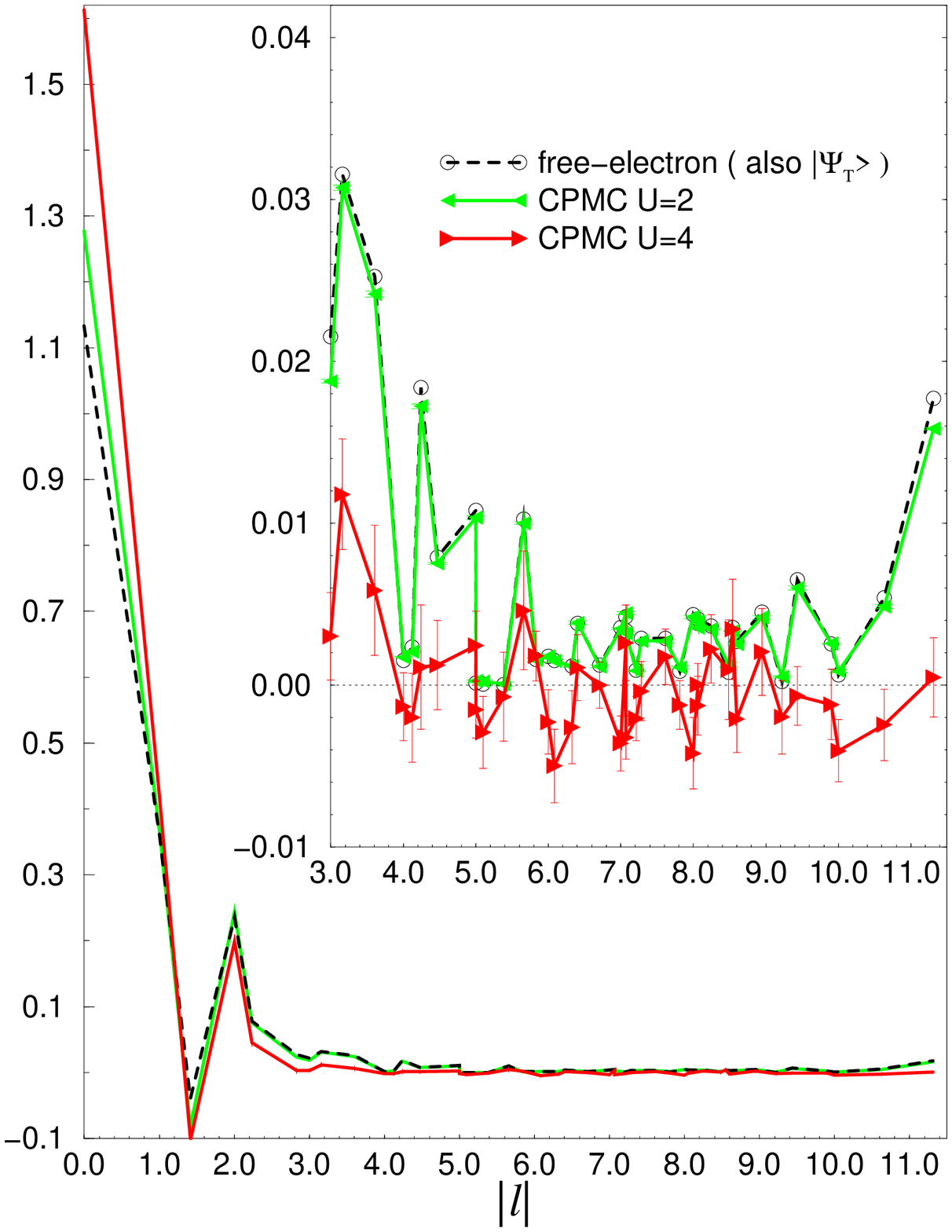}}
\vspace{-0.1in}
\caption{$d_{x^2-y^2}$-wave pairing correlation $D_d({\bf l})$
vs.~distance $l=|{\bf l}|$. TOP: $D_d({\bf l})$
for a 0.85-filled $12\times 12$ lattice at interaction $U=2$ and $4$.
The inset shows the the portion with $l\ge 3$. 
Shown together is the ground-state pairing correlation for the non-interacting system,
whose wave function is also the trial wave function used in the CPMC 
calculations for both values of $U$.
At large distances (inset) $D_d({\bf l})$ systematically
decays as $U$ is increased. Also note that at the irrelevant (short) distances the 
behavior is the opposite.
BOTTOM: Same plot, for a $16\times 16$ system. 
Same trend as that of $12\times 12$
continues, but now the long-range correlation vanishes at $U=4$.}  
\label{figdwave}
\end{figure}

Fig.~\ref{figdwave} shows the $d_{x^2-y^2}$-wave pairing correlations
for $12\times 12$ and $16\times 16$ lattices at $U=2$ and $4$.  The electron
filling is $0.85$, which corresponds to closed-shell cases, with
$N_\uparrow=N_\downarrow=61$ for $12\times 12$ and 
$N_\uparrow=N_\downarrow=109$ for $16\times 16$.  The free-electron wave function
was used as $|\Psi_T\rangle$ in CPMC.  Also shown in each graph is the $D_d({\bf l})$ for
the non-interacting system. We see that the computed
$d_{x^2-y^2}$-wave pairing correlations at both values of $U$ are
bounded from above by the $U=0$ result. Further, consistent with this
observation, we see that at large distances $D_d({\bf l})$ decreases
as $U$ increases from $2$ to $4$. 
In the $16\times 16$ case,
the long-range correlations vanish as the interaction strength is increased to $4$.
The statistics is enough to
discern in this case the irregular oscillations
of $D_d({\bf l})$ around zero. 

Calculation was also done for the $12\times 12$ system at $U=8$ with the same
(free-electron) trial wave function.  The error bars become
substantially larger. $D_d({\bf l})$ oscillates around zero and shows
no indication of enhancement at large distances\cite{pairing}. 

We also note that the behavior of $D_d({\bf l})$ in Fig.~\ref{figdwave}
at short distances is just the opposite of that at large distances,
i.e., $D_d({\bf l})$ increases as $U$ is increased. Due to the large
discrepancy in magnitude between short and long distances, the
integrated value of $D_d({\bf l})$ would therefore not give the
correct trend even at system sizes as large as these.

In summary, the near-neighbor $d_{x^2-y^2}$-wave electron pairing
correlation seems to disappear as the interaction strength increases
and no indication of enhancement is found over the non-interacting
system. Indeed, the behavior of the pairing correlation appears
similar across a fairly broad range of electron fillings, including
the half-filled case, which exhibits pairing correlations of similar
magnitude\cite{pairing}.

We emphasize again that, due to the approximate nature of the CPMC
method, the possibility can not be completely ruled out that the
results were biased such that the pairing correlation was
suppressed. However, the many benchmark calculations, and particularly
the variety of self-consistency checks, indicate strongly that
the results are accurate and robust.

\section{Finite-temperature CPMC algorithm}

Extension of the CPMC algorithm to finite temperatures will allow us
to remove the second set of question marks in Table 1.  

While the standard method of Blankenbecler, Scalapino, and Sugar
(BSS)\cite{BSS,white} has been widely applied, the exponential scaling
has hindered its ability to effectively study true phase
transitions. The difficulty with developing a finite-temperature
counterpart of CPMC lies in the implicit nature of the path-integral
picture in the BSS formalism.  As discussed in Table 1, paths in the
grand-canonical formulation, contrary to ground-state projector QMC,
do not originate or end at a single explicit point in Slater
determinant space. Instead they involve many points, due to the trace
over $N$ (the number of particles, not the number of sites). Indeed
the points do not even have the same ``dimension''\cite{BSS}. A naive
application of the current way of imposing the constrained path
approximation, which would require separate constraining wave functions 
and conditions for paths with
different end points, would therefore not be practical.

We have developed a mechanism to incorporate the constraint into the
analytic evaluation of the trace\cite{finiteT}, which we are currently
testing. In Table 5, we show preliminary results on total energies per
site, for a simple case of the $4\times 4$ system. The chemical
potential is adjusted so that the average filling fraction corresponds
to $5 \uparrow$ and $5 \downarrow$ electrons. As high temperatures,
the new algorithm reproduces the results from AFQMC, while at a low
temperature of $T=0.0625$ ($\beta=16$), it is in good agreement with
the ground-state result from exact diagonalization, as well as that
from the $T=0\,$K CPMC algorithm (cf Table 1).

\begin{table}[htb]
\begin{center}
\caption{Preliminary data from the new finite-temperature algorithm.  The
computed total energies per site, as a function of temperature, are
compared with results from AFQMC\protect\cite{Richard_private} or exact diagonalization.  The
system is $4\times 4$ at $U=4$. The chemical potential is adjusted so
that the average filling fraction corresponds to $5 \uparrow$ and $5
\downarrow$ electrons.
Statistical errors are in the 
last digit and are indicated in parentheses. The $^\star$ indicates
the exact {\em ground-state\/} energy from exact diagonalization.}
\begin{tabular}{clll}
\hline
$T$ & 1 & 0.25 & 0.0625 \\
\hline
CPMC  & -0.8459(2) & -1.1947(6) & -1.2237(6)\\
AFQMC & -0.8457(1) & -1.1947(2) & -1.2238$^\star$\\
\hline
\end{tabular}
\end{center}
\end{table}

\vskip0.3truein

\noindent {\bf Acknowledgements}

\vskip0.1truein

\noindent 
This work was supported by the US National Science Foundation under
grant DMR-9734041 and by an award from Research Corporation.

\end{document}